\documentclass[twocolumn,aps,footinbib]{revtex4}
\usepackage{amsfonts, amssymb}
\usepackage{graphicx, epsfig,bm}
\usepackage{color}

\usepackage{epstopdf}

\newcommand{\be}{\begin{equation}}
\newcommand{\ee}{\end{equation}}
\newcommand{\bea}{\begin{eqnarray}}
\newcommand{\eea}{\end{eqnarray}}

\def\fun#1#2{\lower3.6pt\vbox{\baselineskip0pt\lineskip.9pt
        \ialign{$\mathsurround=0pt#1\hfill##\hfil$\crcr#2\crcr\sim\crcr}}}

\newcommand\lsim{\mathrel{\rlap{\lower4pt\hbox{\hskip1pt$\sim$}}
    \raise1pt\hbox{$<$}}}
\newcommand\gsim{\mathrel{\rlap{\lower4pt\hbox{\hskip1pt$\sim$}}
    \raise1pt\hbox{$>$}}}

\def\dslash{\not{\hbox{\kern-2pt $\partial$}}}
\def\Dslash{\not{\hbox{\kern-4pt $D$}}}
\def\Oslash{\not{\hbox{\kern-4pt $O$}}}
\def\Qslash{\not{\hbox{\kern-4pt $Q$}}}
\def\pslash{\not{\hbox{\kern-2.3pt $p$}}}
\def\kslash{\not{\hbox{\kern-2.3pt $k$}}}
\def\qslash{\not{\hbox{\kern-2.3pt $q$}}}

 \newtoks\slashfraction
 \slashfraction={.13}
 \def\slash#1{\setbox0\hbox{$ #1 $}
 \setbox0\hbox to \the\slashfraction\wd0{\hss \box0}/\box0 }

\def\ee{\end{equation}}
\def\be{\begin{equation}}

\begin{document}
\setlength{\unitlength}{1mm}

\title{Measuring dark energy spatial inhomogeneity with supernova data}
\author{
Asantha Cooray$^1$, Daniel E. Holz$^2$, and Robert Caldwell$^3$}
\affiliation{$^1$Department
of Physics and Astronomy,  University of California, Irvine, CA 92697}
\affiliation{$^2$Theoretical    Division,   Los   Alamos   National
Laboratory,  Los  Alamos, NM  87545}
\affiliation{$^3$Department  of Physics \& Astronomy, Dartmouth College, 6127 Wilder Laboratory, Hanover, NH 03755}

\date{\today}%

\begin{abstract}
The gravitational lensing distortion of distant sources by the
large-scale distribution of matter in the Universe has been extensively studied. In
contrast, very little is known about the effects due to the large-scale
distribution of dark energy. We discuss the use of Type Ia supernovae as
probes of the spatial inhomogeneity and anisotropy of dark energy. 
We show that a shallow, almost all-sky  survey can limit rms dark 
energy fluctuations at the horizon scale down to a fractional
energy density of $\sim 10^{-4}$. 
\end{abstract}
\bigskip
\pacs{PACS number(s): 95.85.Sz 04.80.Nn, 97.10.Vm }

\maketitle

\noindent \emph{Introduction---}Type Ia supernovae (SNe) are thought to
be excellent standard candles, with a calibrated luminosity distance
known to $\sim15\%$~\cite{Kow08}. By using these SNe to measure relative
distances out to cosmological redshifts it is possible to reconstruct the
expansion history of the Universe \cite{Tonryetal03,Rieetal04,Kow08}.
Over the past decade SNe have helped establish that the expansion of the
Universe is
accelerating~\cite{riess-1998,perlmutter-1999,Rie06,Wood,Astier}. Current
and proposed SN surveys aim to significantly expand and refine the
database of SN measurements, and thereby improve our understanding of the
dark energy phenomena thought responsible for the accelerating expansion.

The next generation of SN observations will attempt to answer a
fundamental question about dark energy: whether it is static, indicative
of a cosmological constant, or time-evolving, such as quintessence and a
wide range of alternative theories \cite{Sarkar2}. 
In addition to variations in time, it
is also prudent to consider the possibility of spatial variations.
Fluctuations of the luminosity as a function of direction on the sky,
relative to the global average, may indicate novel physics irrespective
of the dark energy equation-of-state. The aim of this paper is to explore
the degree to which future SN data may place interesting limits on the
homogeneity and isotropy of dark energy. 

Dynamical dark energy is necessarily inhomogeneous; covariance implies
that a time-varying field is equivalently spatially-varying.
Consequently,  fluctuations in dynamical dark energy develop in response
to inhomogeneities in the surrounding radiation and matter fields
\cite{Ma:1999dwa,Dave,Dutta:2006pn}. Where there are fluctuations in
the dark energy density, there should also be spatial variations in the
local expansion rate. These variations may be directly observed as
spatial anisotropy of the luminosity distance-redshift relation. 
In addition, modifications to gravity can masquerade as
dark energy (e.g. \cite{Uzan:2006mf}); spatial fluctuations in the new
gravitational fields can leave a similar imprint on the local expansion
rate and luminosity distances, and there are ongoing efforts to use
fluctuation spectra to test for general departures from General
Relativity
\cite{Hu:2007pj,Daniel:2008et,Jain:2007yk,Bertschinger:2008zb,Flanagan08}.  
Although we do not expect spatial fluctuations in slowly-evolving
quintessence fields, or similarly weak modifications of gravity, to have
measurable impact on the luminosity distance-redshift relation, our lack
of understanding of the physics underlying the dark energy suggests that we
not preclude observable spatial fluctuations.

Dark energy has also been speculatively tied to the breakdown of two
fundamental tenets of cosmology, homogeneity and isotropy. If the
observed cosmic acceleration is due to gravitational effects in a
strongly inhomogeneous universe \cite{Kolb:2005da,Rasanen:2006kp} then we
may expect similar distortions of the expansion rate and luminosity
distance. Anisotropic dark energy
\cite{ArmendarizPicon:2004pm,Battye,Koivisto} may be connected to a
breakdown of global, statistical isotropy, hints of which have been seen
in the CMB \cite{Jaffe:2005pw,Eriksen}. It is conceivable that these
features are also present in the expansion rate of the Universe, and will
be reflected in residuals in the SN luminosity distance
\cite{Koivisto:2007bp}. Furthermore, tests for angular variations in the
luminosity distance can be used to constrain novel photon interactions,
such as photon-to-axion conversion \cite{Csaki,Song:2005af}, which can
distort our view of the cosmic expansion history.

While we do not yet have an understanding of the physical nature
of dark energy, certain candidate models predict spatial
fluctuations in the energy density of dark energy, relative to the global
average, with amplitude $10^{-4}$ on horizon-size length scales
\cite{Dave,Gordon3}. This scale suggests that the best probe of dark
energy clustering may be the CMB, in combination with other measures of large
scale structure. An all-sky CMB map can be used to study dark energy
anisotropies based on differences in the distance to the last scattering surface
in different directions. Unfortunately, the CMB integrates out to $z\sim1100$,
and thus is insensitive to dark energy physics at relatively low redshifts. CMB
limits for dark energy horizon-scale rms fractional fluctuations are not expected to be
better than 10$^{-2}$.

The clustering of dark energy in a supernova sample manifests itself as
correlated fluctuations of the observed luminosity residuals. Of course, 
a variety of conventional gravitational phenomena also distort the
paths of photons from distant SNe~\cite{Durrer,Hui}. The dominant effect
at small angular scales ($\lesssim 5\mbox{ arcmin}$) and high redshift
($z>0.5$) is the lensing magnification due to the intervening mass
fluctuations \cite{Frieman,Holz,Sarkar}. This lensing effect results in
correlated spatial anisotropies in the luminosity distance estimated from
nearby SNe on the sky, and can be measured in terms of magnitude
residuals relative to the best-fit cosmological distance relation at a
given redshift \cite{Cooray1}.  Previous studies have shown that a deep
SN survey can be used to measure 
magnification statistics of the Universe~\cite{Cooray2}.

There are also correlated fluctuations due to Sachs-Wolfe and integrated
Sachs-Wolfe-like effects, consisting of weighted integrals of time
derivatives of the scalar gravitational potential along the line of
sight, but these are expected to be too small to be relevant \cite{Durrer}. 
Finally, there are spatial correlations due to the bulk flows in which the SN host
galaxies are embedded \cite{Cooray3,Neill,Gordon2}, although these are
only important at lower redshifts ($z<0.2$). Efforts to discover
correlated luminosity fluctuations induced by dark energy must overcome
these nuisance effects.

\smallskip
\noindent \emph{Calculational Method---} We calculate the fluctuations in
the apparent luminosity due to perturbations in the dark energy
component \footnote{It is also possible to perform our
analysis in luminosity distance instead of luminosity, with identical
results.}. We also incorporate the effects due to inhomogeneities in the matter, which
account for conventional gravitational lensing.
We lack a theoretically compelling model for the dark energy,
and thus confine ourselves to a purely phenomenological description of
dark energy inhomogeneities, motivating a model-independent measurement
that can be used to guide future tests of specific models of clustering.

We model the luminosity distance, $d_L(z, \hat{\bf n})$, to a source at
redshift $z$ as a function of sky location $\hat{\bf n}$.  In a general Robertson-Walker space-time, the
luminosity distance along a given line-of-sight is given by
\begin{eqnarray}
d_L(z,\hat{\bf n}) &=& \frac{(1+z) c}{H_0\sqrt{\Omega_K}} \sinh(\sqrt{\Omega_K}\int_0^z dz'/ [\Omega_{x}(z',\hat{\bf n}) \cr
&+&   \Omega_M(1+z')^3+ \Omega_K(1+z')^2]^{1/2}),
\end{eqnarray}
where $c$ is the speed of light, $H_0$ is the Hubble constant today,
and $\Omega_M$, $\Omega_K$, and $\Omega_{x}$ are the matter, curvature, and
dark energy density in units of the critical density. We specialize to the spatially-flat case
for this study. We note that for
the line-of-sight integral, the relevant densities are the actual
densities in the given direction (including the effect of
spatial variations), not the globally-averaged densities. The apparent
luminosity of a source is given by $L(z, \hat{\bf n}) = \bar L \times
(\bar{d_L}/d_L(z, \hat{\bf n}))^2$ where $\bar L$ and $\bar{d_L}$ are the
mean  luminosity and luminosity distance, respectively, defined
by the underlying, unperturbed cosmology~\cite{Holz}. Perturbations are
given by $\delta_{L}(z, \hat{\bf n}) = [L(z, \hat{\bf n}) -
\bar{L}]/\bar{L}$. We build a phenomenological model for the dark energy
inhomogeneities, characterizing the fluctuations as $\delta_x(z,\hat{\bf
n})\equiv[\Omega_x(z,\hat{\bf n})-\bar{\Omega}_x(z)]/\bar{\Omega}_x(z)$,
relative to the mean dark energy density, $\bar{\Omega}_x$, at each
redshift. We can then write the three dimensional power spectrum for the
dark energy fluctuations in Fourier space as $\langle \delta_x({\bf
k})\delta_x({\bf k'})\rangle  = (2\pi)^2 \delta({\bf k}+{\bf k'})
P_x(k)$. By observing these perturbations as a function of angular scale
and redshift, we can attempt to quantify the underlying inhomogeneity of
the dark energy \cite{Mortsell}.

In the limit of small number statistics, as would be the case with
current and next-generation large SN surveys, it may be more practical to
measure the angular power spectrum of luminosity fluctuations. Assuming
statistical isotropy, the angular power spectrum is given by $\langle
(\delta_L)_{\ell m}^\star (\delta_L)_{\ell' m'}\rangle = C_\ell^{LL}
\delta_{\ell \ell'} \delta_{m m'}$, where $(\delta_L)_{\ell m}$ are the
multipole moments of the luminosity fluctuations. The expected angular
power spectrum of (fractional) luminosity fluctuations, $\delta_L$, for a
sample of SNe at redshift $z$, is given by:
\begin{equation}
C_\ell^{LL} = \int dr  \frac{W^2(r)}{d_A^2(z)}
P_{x}\left(k=\frac{\ell}{d_A},r\right),
\label{eq:cl}
\end{equation}
where
\begin{equation}
W(r)= 3\int dr' n(r'){\Omega_m H_0^2\over c^2 a(r)}{d_A(r) d_A(r' - r)\over d_A(r')}.
\end{equation}
Here $d_A(z)$ is the comoving angular diameter distance, $r$ is the
comoving radial distance, and $n(r)$ is the radial distribution of SNe.

Equation 2 neglects the contributions of gravitational lensing due to
matter inhomogeneities. In what follows we will consider lensing as a
source of noise, such that any measurement of $P_x(k)$ above this noise
represents a measurement of the dark energy inhomogeneity power
spectrum. Instead of a specific model \cite{Takada}, 
we characterize the power spectrum of dark energy fluctuations
with a phenomenological model in Fourier space:
\begin{equation}
\frac{k^3P_{x}(k,z)}{2\pi^2} = \delta_{xH}^2 \left(\frac{ck}{H_0}\right)^{n_x+3}
e^{-k/k_c}(1+z)^{-s}\, ,
\label{eq:model}
\end{equation}
where $\delta_{xH}$ is the amplitude at horizon scales, $n_x$ is the
tilt to the spectrum, $s$ is the exponent in the growth factor, and
$k_c$ is the cut-off scale for fluctuations. For
quintessence fields the cut-off wavenumber is set by the dynamics of the
equation of state. Formally, $P_x$ is
the power spectrum of an additional clustering component to
potential fluctuations, 
which we evaluate in the conformal-Newtonian/longitudinal gauge 
(we absorb a prefactor proportional to $\Omega_x$ to $\delta_{xH}$ 
\cite{Mota2}). 
The parameters ($\delta_{xH}$, $n_x$, $s$, $k_c$) can be constrained with future SN data.  Note that while we
have written a power spectrum associated with dark energy fluctuations, the same
power spectrum may also describe the impact of modifications to gravity, or
exotic physics of the dark sector.

Gravitational lensing due to the inhomogeneous matter distribution sets a
lower limit on the strength of detectable dark energy fluctuations. The
dark energy signal must rise above this  amplitude, and thus the
$1\sigma$ upper limit on $\delta_{xH}$ (given other parameters) of a {\em null}\/ experiment is given by
\begin{equation}
\frac{1}{\left(\delta_{xH}^{\rm min}\right)^2} = 
\sum_{l=l_{\rm min}}^{l_{\rm max}} 
\left(\frac{\partial C_\ell^{LL}}{\partial \delta_{xH}}\right)^2_{\delta_{xH}=0} 
\frac{1}{\left(\Delta C_\ell^{LL}\right)^2} \, ,
\label{eqn:ainv}
\end{equation}
where
\begin{equation}
\Delta C_\ell^{LL} = \sqrt{\frac{2}{(2\ell+1)f_{\rm sky}\Delta \ell}} 
\left[ C_\ell^{\mu} + \frac{\sigma^2_L}{N_{\rm SN}}\right] \,.
\label{eqn:bin}
\end{equation}
Here $C_\ell^{\mu}$ is the anisotropy power spectrum of SN luminosities
expected from cosmic magnification due to mass fluctuations
\cite{Cooray2}, $N_{\rm SN}$ is the surface density of SNe (number per
steradian), $\sigma_L$ is the uncertainty in the luminosity measurement
from each SN, $f_{\rm sky}$ is the fraction of sky covered by the
survey,  and $\Delta\ell$ is the binning width in multipole space. When
considering the extent to which finite SN surveys can search for
inhomogeneities, we take $l_{\rm min}= 180^\circ/\theta$ where $\theta$ is
the one-sided area of the survey. We assume an uncertainty in
the (calibrated) intrinsic luminosity of SNe given by $\sigma_L=0.1$.

\begin{figure}[!t]
\epsfxsize=3.5in
\centerline{
\epsfbox{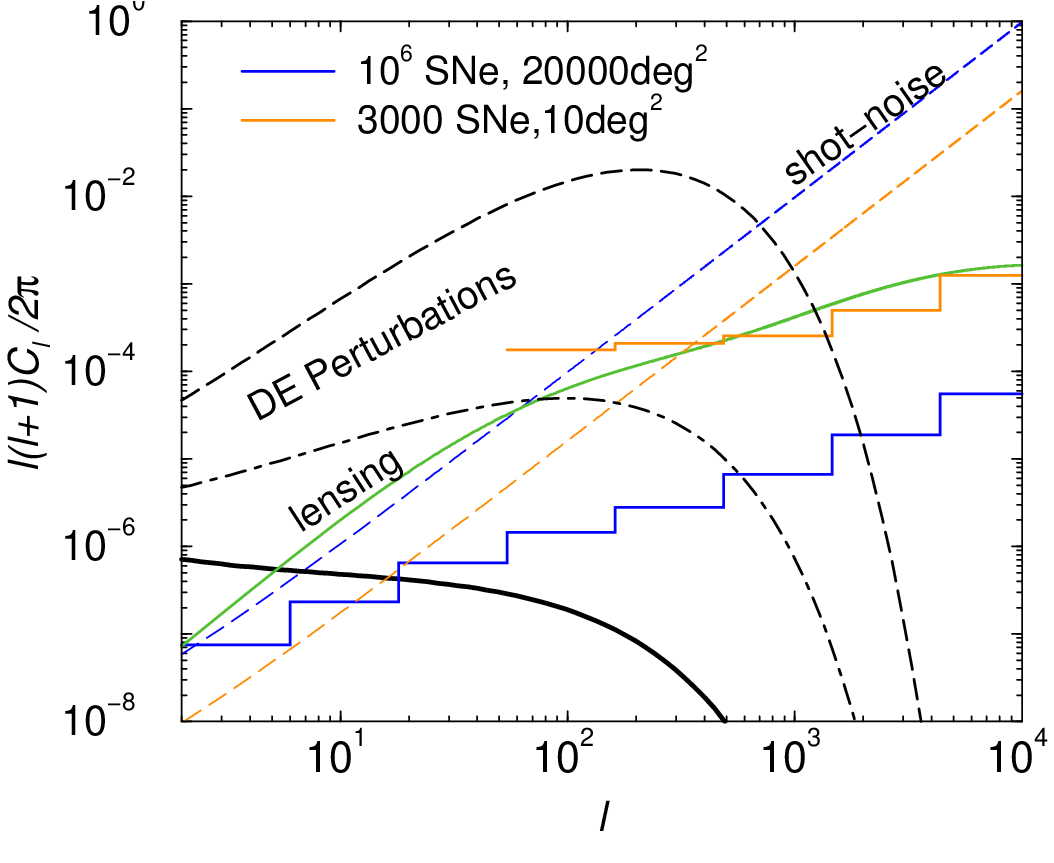}}
\caption{Angular power spectrum ($C_\ell^{LL}$) of SNe luminosity, relative
to the mean, as a function of redshift. The thin solid ``lensing'' line is the power
spectrum due to gravitational lensing: the luminosity fluctuations generated by
magnification associated with foreground mass perturbations. We show three examples of power
spectra due to large-scale dark energy inhomogeneities, characterized by (see
eq.~(\ref{eq:model})) $\delta_{xH}^2=10^{-5}$, $k_c=0.1$ $h$ Mpc$^{-1}$, $s=2$ and 
 $n_x=-2$ (bottom solid line), -1 (middle dot-dashed line), and 0 (top dashed line).
The two dashed ``shot-noise'' curves are  for surveys with 3,000 SNe in 10 deg$^2$ (bottom) and 10$^6$ SNe in
20,000 deg$^2$ (top). The two stair-step 
lines show the total noise in measuring
luminosity anisotropies (equation~6), including the contribution from
lensing  term $C_l^\mu$. The width of the step is
equal to the bin width $\Delta \ell$.
}
\label{figure1}
\end{figure}

\smallskip
\noindent \emph{Results---} Fig.~\ref{figure1} summarizes our results,
illustrating the sensitivity of future SN surveys to clustering of the
dark energy. We show the angular clustering power spectrum of luminosity
variations relative to the mean luminosity. The weak lensing power
spectrum (thin solid curve) dominates at arcminute angular scales and at high
redshifts~\cite{Cooray2}. We also show examples of dark energy
clustering, arising from models with $\delta_{xH}^2=10^{-5}$, $s=2$, 
$k_c=10^{-1}$ $h$ Mpc$^{-1}$  for three values of $n_x$. 
These spectra are chosen such that fluctuations in the luminosity residuals become important at
large angular scales, and thus dominate over the perturbations due to
matter fluctuations (gravitational lensing). For comparison,
the amplitude $\delta_{xH}^2$ for horizon-scale perturbations, 
relative to the dark energy density, is of order $\sim 10^{-8}$ for certain dark energy models
 \cite{Dave,Gordon3}. 

For comparison, in Fig.~\ref{figure1} we also plot the shot-noise for two
potential SN surveys (dashed lines): 3,000 SNe in 10
deg$^2$, as can be achieved with deep repeated observations from space, and 
10$^6$ SNe in 20,000 deg$^2$, from a shallow, but wide, SN survey.
The latter is equivalent to what can be achieved with the Large
Synoptic Survey Telescope (LSST; \cite{LSST}). We combine the effects of
lensing and shot-noise, generating a total noise estimate in wide
multipole bins, following equation~(\ref{eqn:bin}). The stair-step curves
show this binned noise for the two SN samples we consider, where we take
$\Delta \ell=2\ell$ for the bin sizes.  

Using the procedure outlined in equations~(\ref{eqn:ainv}) and
(\ref{eqn:bin}), we have estimated the minimum rms amplitude of dark energy
spatial fluctuations at horizon scales that can be detected with 
future SN samples. A survey of 3,000 SNe in 10 deg$^2$ (e.g., from a
space-based {\it JDEM} \cite{DETF}) provides limited constraints due to
the narrow sky coverage, finding limits of $\delta_{xH} <0.2$ if $n_x\sim-3$, and $\delta_{xH} < 10^{-4}$ if $n_x\sim0$.
A shallow, wide survey of 10$^6$ SNe over 20,000 deg$^2$ (e.g., as expected from LSST)
could potentially limit $\delta_{xH} < 3\times10^{-3}$ if $n_x \sim -3$, and 
$\delta_{xH} < 2\times10^{-5}$ if $n_x \sim 0$. These limits are for a growth 
exponent of $s=2$ in
equation~(\ref{eq:model}). 
If dark energy fluctuations are redshift-independent ($s=0$),
then the limits improve by a factor of 2 to 3. 
Independent of $n_x$ and $s$, 
LSST could limit the dark energy quadrupole ($l=2$) to a fractional energy
density as low as $2 \times 10^{-4}$.

The phenomenological dark energy fluctuation power spectrum may be
observationally  determined by mapping the luminosity residuals as a
function of SN redshift, and then measuring the angular power spectrum of
the residuals (i.e., in the same manner that CMB residuals are treated).
The imprint of dark energy perturbations on the CMB, and
cross-correlations with other tracers of large scale structure, is
limited by cosmic variance. By using a shallow, all-sky SN survey the dark
energy perturbation signal is dramatically improved. 
Thus, beyond the CMB, it is likely that
SN data will become the most powerful probe of anisotropic expansion and
dark energy inhomogeneity in the universe.

The multipole moments of the luminosity residuals, $(\delta_L)_{\ell m}$,
can be used for cross-correlation studies with other trace fields of
the large-scale structure. The moments can be used for additional
statistical studies to constrain signatures of global anisotropies, which
may be due to anisotropic dark energy or other direction-dependent
phenomena. In a similar fashion, such studies will benefit from
an all-sky SN sample, which could be obtained with a
shallow SN survey out to $z \sim 1$ rather than a deep (but narrower) survey that
attempts to recover SNe at $z \sim 2$.
Alternatively, the absence of SN luminosity 
fluctuations due to dark energy inhomogeneities consists of a 
null test of the cosmological constant.

The cosmological community is aggressively developing and perfecting
probes of the time dependence of the dark energy equation-of-state. It is
to be noted that time-dependent models of the dark energy also necessarily predict
spatial dependences.  Type Ia supernovae have become one of the most
powerful probes of the dark energy equation of state, and are expected to
provide precision constraints on possible time variation. We show that 
these same supernovae also offer one of the most powerful methods to constrain spatially
inhomogeneous models of the dark energy.

We have presented a simple phenomenological model for large-scale spatial
fluctuations of the dark energy, and have derived limits to how well this model
can be constrained with various proposed surveys. It will be useful to quantify the
effective variations in dark energy for a variety of theoretical models.
In addition, it would be worthwhile to consider other potential
cosmological probes of dark energy inhomogeneity, including signatures in
weak lensing shear measurements of galaxy shapes, as well as spatial
inhomogeneities in baryon acoustic oscillation measurements. The advent
of almost all-sky SN samples, such as would be expected from LSST, will
sharply constrain the amount of spatial variation in the dark energy.

\smallskip
This work  was supported by LANL IGPP Astro-1603-07 (DEH \& AC), NSF CAREER
AST-0645427 (AC), and AST-0349213 (RRC). AC and DEH thank the Aspen
Center for Physics for hospitality while this paper was initiated. AC and RRC thank Caltech
for hospitality while this paper was completed.


\end{document}